\documentclass[12pt]{article}

\usepackage{amsmath}
\usepackage[english]{babel}
\usepackage{amssymb}
\usepackage{graphicx}
\usepackage{cite}
\usepackage{subfigure}
\def\beq{\begin{equation}}
\def\eeq{\end{equation}}
\def\bea{\begin{eqnarray}}
\def\eea{\end{eqnarray}}
\def\nn{\nonumber}

\def\nn{\nonumber}

\begin{document}

\title{(Pseudo)-Dirac neutrinos and leptogenesis\footnote{Talk given at the Dark Side of the Universe
Conference 2006 (DSU2006), Madrid, Spain, 20-24 June 2006}}

\author{Steven Abel and V\'eronique Pag\'e \\
    {\em Centre for Particle Theory,} \\
{\em University of Durham, Durham, DH1 3LE, UK}}

\maketitle

\abstract
\noindent We discuss how Dirac neutrinos can naturally be generated in supersymmetry and 
how they allow for an Affleck-Dine leptogenesis scenario, in which a left-right 
asymmetry is generated in the sneutrino sector, 
the left part of which is transferred to a
baryon asymmetry via sphaleron transitions. No exotic fields need 
to be added to the MSSM other than the right-handed neutrino.

\section{Introduction}
As the nature of neutrinos still eludes us, it is worth bearing in mind
all the alternative possibilities for neutrino masses. In particular 
the Majorana or Dirac nature of neutrinos has not yet been established and, in spite
of (or perhaps because of) the current hegemony of the see-saw mechanism,
it is interesting to ask whether the latter scenario is a reasonable possibility.
This talk is based on two papers that demonstrate the naturalness 
of Dirac neutrinos in the context of supergravity, paying particular 
attention to their masses  \cite{abelsakis} and to baryogenesis \cite{abelpage}.
These are the two issues that need to be addressed if 
Dirac or pseudo-Dirac neutrinos are to be considered a viable alternative 
to the more usual see-saw mechanism. 

Let us begin by recalling the basics; considering only one neutrino flavour 
for simplicity, the most general neutrino mass matrix can be written as
\bea
- \mathcal{L}_{mass} = \frac{1}{2}\overline{\chi^c_L} M_\nu \chi_L + \frac{1}{2}\overline{\chi_L} M_\nu \chi_L
\eea
with 
\bea
\chi = \begin{pmatrix} \nu \\ \nu^c \end{pmatrix}, & &\quad \overline{\chi^c}=-\left( \overline{\nu^c},\overline{\nu} \right), \nn \\
M_\nu=& &\begin{pmatrix} m_L & m_D \\ m_D^T & M_N \end{pmatrix} ~.
\eea
The various scenarios for the neutrino mass are essentially differentiated 
by the importance of each term in the mass matrix $M_\nu$.  In 
the see-saw scenario \cite{minkowskiseesaw,ssmgellmann,ssmyanagida,ssmmohapatra,weinbergseesaw,valleseesaw1,lazaridesseesaw,valleseesaw2,valle_neutrino}, 
the mass terms are such that
\bea
M_{\nu,ss}=\begin{pmatrix} m_L & \ll & m_D \\ m_D^T & \ll & M_N \end{pmatrix} ~.
\eea
There is little 
oscillation between active and sterile because of small mixing angle.
The 'sterile' and 'active' states $\nu^c$ are very nearly mass eigenstates 
and, when giving it a Dirac mass $m_D$ similar to other fermions, the active 
state obtains naturally a very small mass $m_D^2/M_N$.  
The origin of the mass 
scale $M_N$, however, is left unexplained. At the opposite extreme, the pure Dirac 
neutrino has a mass matrix
\bea
M_{\nu,d}=\begin{pmatrix} 0 & m_D \\ m_D^T &  0 \end{pmatrix} ~,
\eea
and there is no oscillation between active and sterile because of mass degeneracy. 
Here there is no new energy scale to be explained, apart from, obviously, the 
neutrino mass scale itself.  This has been often presented as a major drawback 
for the Dirac neutrino scenario as opposed to the see-saw one, however 
a natural explanation {\em does} exist in the context of supergravity, related
to the solution of the $\mu$-problem by Giudice and Masiero \cite{giudice}. 
The $\mu $ problem is to explain why the mass term for the higgs
fields $\mu H_u H_d $ should have a value similar to the supersymmetry 
breaking scale, to which it is apparently unrelated. The explanation 
of ref.\cite{giudice} is that the $\mu$-term 
does not appear at first order in the 
superpotential, but does appear in the K\"ahler potential of supergravity. 
Thus if 
\beq
K \supset H_u H_d + H.c.\, ,
\eeq
then a term $\mu H_u H_d $ 
is communicated to the effective potential by gravity in the 
same way as the supersymmetry breaking, and hence
 with $\mu \sim M_{SUSY} \sim $ 1 TeV.

Could such an explanation work for Dirac neutrino masses as well?
The numbers certainly suggest that it could, as has 
been occasionally noted in the literature \cite{arkanihamed,borzumatinomura,
casasetal,kitano,arnowittetal}.  Consider for instance supergravity with 
K\"ahler potential $K$ and superpotential $W$; let us remove the 
neutrino mass terms from the superpotential and place them 
instead in the K\"ahler potential, in the form
\beq
K \supset \frac{LH_u\bar{N}}{M} + \frac{LH_d^*\bar{N}}{M} + H.c.
\eeq
with $\bar{N}$ being the RH neutrino and $M$ the scale at which higher 
dimensional operators make their appearance in $K$.  Assume now 
that SUSY breaking is communicated by gravity, with $m_{3/2} \sim 1$TeV, and 
for the sake of the argument, consider $M \approx M_P$.  Then, taking 
$\left< H_u \right>  \approx m_{top}$, we obtain
\beq
m_{\nu} \approx \frac{m_{3/2}}{M}\left( \left< H_u \right> +\left< H_d \right> \right) \sim 10^{-4} \mathrm{eV} ~,
\eeq
which is intriguingly close to the measured value of $0.04-0.05$eV, assuming that 
the mass-squared differences are indicative of the actual masses.  A detailed 
calculation has been made in \cite{abelsakis}, along with the suggestion of 
an R-symmetric model that could accomodate such a scenario.  Here we shall 
simply insist on the fact that the inclusion of the neutrino mass-scale 
'problem' within the framework of Susy and Sugra allows for the appearance 
of scenarios that do not require the addition of a new mass scale.  

\section{Affleck-Dine Dirac leptogenesis} \label{section: AD}
Let us now turn to the question of the leptogenesis mechanism. The 
original leptogenesis scenario 
\cite{leptogen} requires the presence of a Majorana mass and its accompanying 
mass scale.  A leptogenesis scenario in the absence of a Majorana 
masse, sometimes called \emph{neutrinogenesis}, was
introduced by Dick \& al \cite{dicketal}. That model was presented
in the context of a modified SM  with an 
additional heavy Higgs-like doublet.  The main feature was 
the smallness of the neutrino Yukawas which effectively
hides a leptonic asymmetry away from the sphalerons, which are blind to 
the RH sector.  Various implementations of this scenario have been suggested \cite{sakislepto,thomastoharia}.  Here we point out that, if we are willing to include the question 
of Dirac leptogenesis within the framework of SUSY, the Affleck-Dine (AD)
mechanism \cite{admech} allows a very efficient 
implemententation of neutrinogenesis in just the MSSM + right-handed neutrinos.  
The AD mechanism allows the production of a $(\tilde{\nu}_L-\tilde{\nu}_R)$ 
current and, although lepton-number is conserved, only the LH lepton number 
can be converted to a baryon number through sphalerons, while the RH
component is hidden by the smallness of the Yukawa coupling \cite{abelpage}.  

Let us first review the original neutrinogenesis scenario and the model 
suggested by \cite{dicketal}.  The model suggested requires the 
addition of an heavy, Higgs-like doublet.  Let us call this 
doublet $\phi$; by the usual 'drift and decay' mechanism, the decay of $\phi$ 
starts a chain of reactions that leads to the desired baryon asymmetry:
\beq
\phi \to \nu_L + \bar{\nu_R} \to -\alpha(B+L) +\nu_L + \bar{\nu_R} ~.
\eeq
Because of the smallness of the neutrino mass $\nu_R$ is inert, yet it still 
holds a lepton number.  As usual, sphalerons transfer (only the LH) lepton 
number to baryons.  CP violation is provided by the decay of $\phi$.  
Our proposal is that the AD mechanism \cite{admech} can play the 
role of the $\phi$ decay, thus rendering the addition of the $\phi$ field 
unnecessary.  In the AD mechanism, scalar fields 'slow-roll' along flat 
directions of the superpotential, which causes them to accumulate some quantum 
number.  The original AD scenario saw the fields accumulating a $B$-number 
directly.  Here they will instead accumulate a 'left-right' (LR) asymmetry, 
allowing neutrinogenesis to create the baryon number of the Universe.

\begin{center}
\begin{figure}
\begin{tabular}{r l}
\begin{minipage}{0.7\textwidth} 
\includegraphics[width=0.7\textwidth]{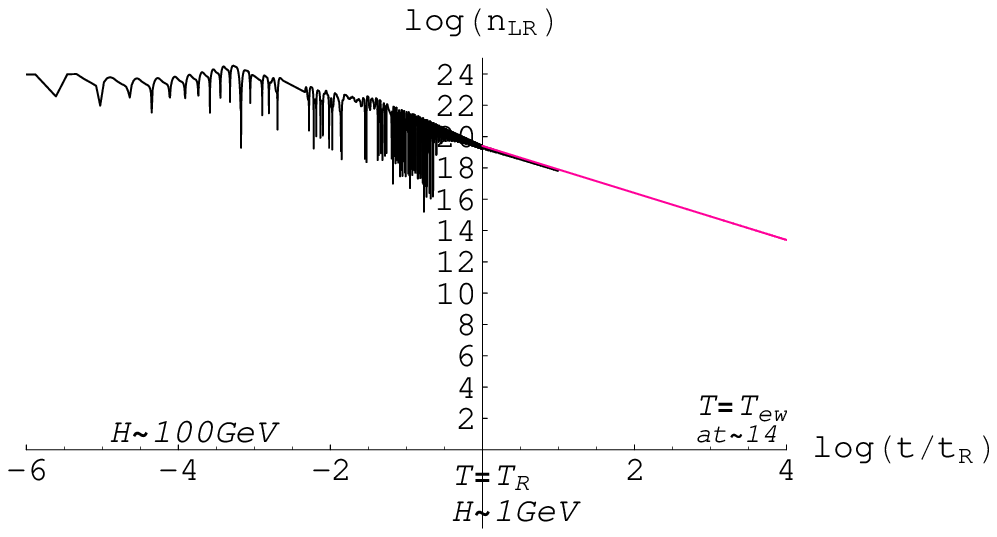}
\end{minipage}
\begin{tabular}{l}
\begin{minipage}{0.3\textwidth} 
\includegraphics[width=0.7\textwidth]{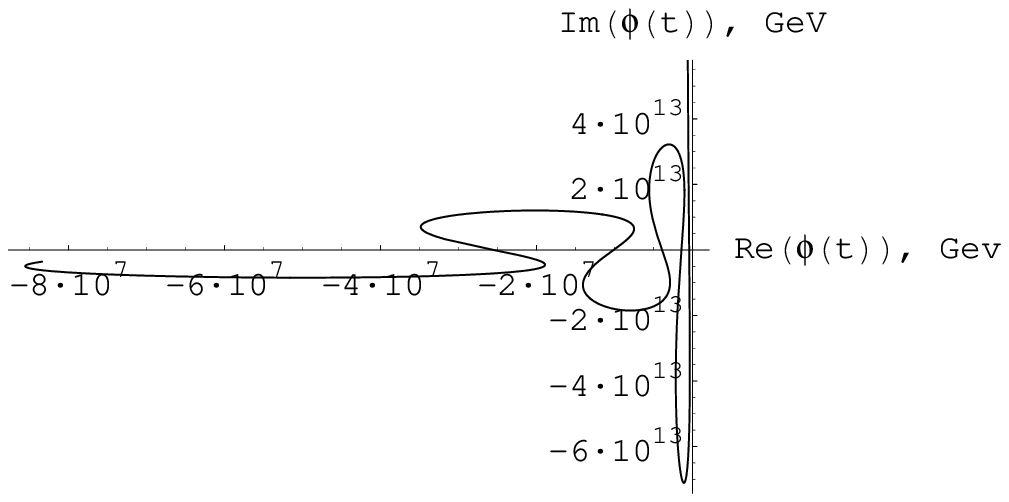}  
\includegraphics[width=0.7\textwidth]{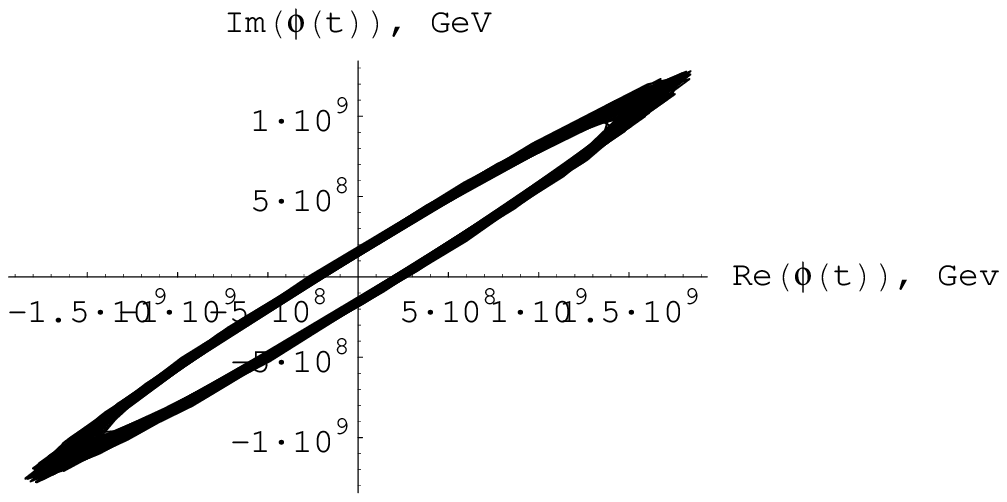}
\end{minipage}
\end{tabular}
\end{tabular}
\caption{Time evolution of the generated LR asymmetry for typical parameters.  
The behaviour of the $\phi$ field is also shown for early (shortly 
before $H \sim 100$GeV) and late (post-reheating) times.}
\label{fig: evolution}
\end{figure}
\end{center}

Consider the superpotential of the \emph{effective} global Susy theory:
\beq
W =  Y_UQH_uU^c + Y_DQH_dD^c + Y_ELH_dE^c  + Y_{\nu}LH_uN^c + \mu H_uH_d ~.
\eeq
It possesses two $D$-flat directions, $LH_u$ and $N^c$:
\bea
\begin{array}{ccccccccc}
L &=& \frac{1}{\sqrt{2}} \left( \begin{array}{c} \phi \\ 0 \end{array} \right),& \qquad H_u &=& \frac{1}{\sqrt{2}} \left( \begin{array}{c} 0 \\ \phi \end{array} \right),& \qquad N^c  &=& \bar{\widetilde{\nu}} ~.
\end{array}
\eea
We should note that these directions are not perfectly $F$-flat due to the 
presence of the Yukawa couplings;  however these are small enough not to 
endanger the success of the mechanism, and indeed it is the non-zero $F$-terms that 
will indirectly determine the baryon number.  The scalar potential in which these 
fields are evolving is given by:
\bea
V &=& V_{SB}+V_{Hubble}+V_F \nn\\
&=& \left(m_{\phi}^2-c_{\phi}H^2  \right)|\phi|^2 + \left(m_{\nu}^2-c_{\nu}H^2  \right)|\bar{\widetilde{\nu}}|^2 \nn\\
&&+ \left( Y_{\nu}(A+c_AH)\phi^2 \bar{\widetilde{\nu}} +h.c. \right) + \frac{|Y_{\nu}|^2}{4} |\phi^2|^2 + |Y_{\nu}|^2|\bar{\widetilde{\nu}} \phi|^2
\eea
where $ V_{SB}$, $V_{Hubble}$ and $V_F$ are  the SUSY-breaking, 'Hubble' and $F$-term 
potentials, respectively \footnote{The 'Hubble' potential is the effective potential 
due to SUSY breaking in the early Universe.  It is the presence of these terms 
that allow the flat directions to develop large expectation values during inflation - 
see 
\cite{drandallt1,drandallt2}}.  We can see now how the LR asymmetry:
\beq
n_{LR}=n_L-n_R
\eeq
where
\bea
\begin{array}{cccccc}
n_L&=&\frac{\imath}{2}\left( \dot{\phi}^*\phi - \dot{\phi}\phi^* \right),& \qquad n_R &=&-\imath\left( \dot{\bar{\widetilde{\nu}}}^*\bar{\widetilde{\nu}}-\dot{\bar{\widetilde{\nu}}}\bar{\widetilde{\nu}}^* \right) 
\end{array}
\eea
has a non-trivial evolution.  Indeed the evolution of $n_{LR}$ is obtained through 
solving the equation of motion for each field, of the type:
\beq
\ddot{\phi}+ 3H\dot{\phi}+ \frac{\partial V}{\partial \phi}\phi~.
\eeq
Doing this, we obtain:
\beq
\dot{n_{LR}}+3Hn_{LR}=4\mathrm{Im}\left( Y_{\nu} A \phi^2 \bar{\widetilde{\nu}} \right) ~.
\eeq
The behaviour of the asymmetry is given in Figure (\ref{fig: evolution}), along 
with the evolution of the scalar field $\phi$; the approximate analytical behaviour 
for each phases of evolution is also given in Table (\ref{table: evolution}).  
We should mention that it is obviously necessary that the $\bar{\widetilde{\nu}}$ 
oscillations decay after the electroweak phase transition.  With the lifetime 
given by $\tau_{decay}=4 \pi / (Y_{\nu}^2m_{\tilde{\nu}}) $ we see that indeed 
$T_{decay}\sim 100$MeV.  The LH sneutrinos, however, will quickly decay to 
neutrinos (this is instantaneous; see \cite{abelpage}), and this brings us back 
to original neutrinogenesis.  Sphalerons can transfer this (LH) lepton number 
to baryons, although at all times the LH lepton number is accompanied by an equal 
and opposite lepton number hidden (to sphalerons) in the inert RH sneutrinos.  
The fact that the RH, sterile sneutrinos hold an asymmetry equal and 
opposite to the baryonic one opens the interesting possibility that they 
form the cold component of dark matter.

\begin{table}[t]
\begin{tabular}{|c|c|c|c|}
\hline
$H>>m_{3/2}$& $R \sim t^{2/3}$ & chaotic motion& $R^3H^2\phi^2=const \Rightarrow n_{LR} \sim const$ \\
\hline
$H(T_R)<H< m_{3/2}$ & $R \sim t^{2/3}$ & cyclic motion & $R^3m_{\phi}^2\phi^2 = const \Rightarrow n_{LR} \sim t^{-1}$ \\
\hline 
$H<H(T_R)$ &  $R \sim t^{1/2}$& cyclic motion &  $R^3m_{\phi}^2\phi^2 = const \Rightarrow n_{LR} \sim t^{-3/2}$ \\
\hline 
\end{tabular}
\caption{Approximate analytical behaviour of the LR asymmetry.  The early chaotic 
period is typical of the flat directions used here, as those are not perfectly $F$-flat.}
\label{table: evolution}
\end{table}

\section{The baryonic asymmetry} \label{section: asymmetry}
To be able to approximate the LR asymmetry, let us recall that before reheating, 
while the inflaton oscillations dominate the energy of the Universe, both the 
inflaton and the asymmetry behave like matter.  Thus we can use the inflaton 
energy density to keep track of the asymmetry:
\beq
\frac{\rho_{n_{LR}}}{\rho_I} \sim \frac{m_{3/2}^2 \left| A/Y_{\nu} \right|^2}{m_{3/2}^2 M_P^2} ~.
\eeq
After reheating, it is the ratio of the asymmetry with entropy that remains 
constant, and now $\rho_{n_{LR}}=m_{\phi, \bar{\widetilde{\nu}}}n_{LR}$.  Hence 
the asymmetry is: 
\bea
n_{LR} &\approx& \frac{|A/Y_{\nu}|^2}{M_P^2}\frac{T_R}{m_{\phi}} = 10^{-9}\left| \frac{A}{100 \mathrm{GeV}} \right|^2 \left| \frac{10^{-12}}{Y_{\nu}} \right|^2 \left| \frac{T_R}{1 \mathrm{TeV}} \right|  \left|\frac{100 \mathrm{GeV}}{m_{\phi}} \right| ~.
\eea
This asymmetry is related to the baryon number in a fairly straightforward way 
\cite{harveyturner,dreinerross}:
\bea
\begin{array}{cccccccc}
B &= & L & = &  \frac{8}{23}n_{L}^{(R)}  &  \qquad T > T_{ew}, ~.
\end{array} 
\eea
Hence the observed baryon number is obtained for reheating temperatures 
of order $1$TeV.  Should the RH sneutrinos be the dark matter, this last relation 
allows us to constrain their mass via 
\beq
m_{DM} = \frac{8}{23}\frac{\Omega_{DM}}{\Omega_{b}}m_b\, .
\eeq

\section{Conclusion}
The possibility of neutrinos being Dirac or pseudo-Dirac particles is 
still very much alive.  First, within the framework of supergravity, neutrino 
masses can be made naturally small.  Their scale can be related to the scale 
hierarchy between the weak and Planck scale, in much the same way  as for 
the $\mu$-term.  No new scale is necessary to explain the neutrino mass.  
Moreover, still within SUSY, leptogenesis with Dirac neutrinos can be 
easily implemented using the Affleck-Dine mechanism.  With a reheating 
temperature of order $1$TeV, the right order of magnitude for the baryon number 
of the Universe is obtained.  This requires no new fields to be added to the 
MSSM beyond the right handed neutrino. In the case of pure Dirac neutrinos
the $B-L$ number of the visible sector is connected to an equal and opposite 
right-handed sneutrino number, and  
this provides an intriguing connection between the dark matter density and the 
baryon number of the Universe. This link has been obtained in the past in 
various works \cite{dmbarr1,dmbarr2,dmkaplan,dmthomas,dmkuzmin,dmkusenko,dmfarrar,
dmhooper,dmkitano1,tytgat,dmkitano2}, 
but here we see it arising quite naturally. 

\bibliographystyle{unsrt}
\bibliography{madrid_proceeds_v4}

\end{document}